# The quantum hydrodynamic formulation of Dirac equation and its generalized stochastic and non-linear analogs

Piero Chiarelli

*National Council of Research of Italy, Area of Pisa, 56124 Pisa, Moruzzi 1, Italy*
*and*
*Interdepartmental Center "E.Piaggio" University of Pisa*
Phone: +39-050-315-2359
Fax: +39-050-315-2166

Email: pchiare@ifc.cnr.it.

Abstract: The quantum hydrodynamic-like equations as a function of two real sets of variables (i.e., the 4x4 action matrix and the 4-dimensional wave function modulus vector) of the Dirac equation are derived in the present work. The paper shows that in the low velocity limit the equations lead to the hydrodynamic representation of the Pauli's equation for charged particle with spin given by Janossy [1] and by Bialynicki et al [2]. The Lorentz invariance of the relativistic quantum potential that generates the non-local behavior of the quantum mechanics is discussed.

# 1. Introduction

In the present paper the author develops the hydrodynamic formulation of the Dirac equation.
The quantum hydrodynamic analogy (QHA) describes more clearly the origin of the non-local quantum character deriving by the quantization condition [2] and it results useful in treating problems at the edge between the quantum and the classical regime.
In the hydrodynamic quantum equations (HQEs) [3] the non-local restrictions come by applying the quantization of vortices [2] and by the elastic-like energy arising by the quantum pseudo-potential but not from boundary conditions.
In the low speed limit the Schrödinger equation is a differential equation where the non-local character of evolution is introduced by the initial and boundary conditions that must be defined for describing the physical problem.
In the case of charge particles, the non-local properties of the Schrödinger equation come also from the presence of the electromagnetic (em) potentials that depend by the intensities of em fields in a non-local way (e.g., Aharonov –Bohm effect).
In the corresponding hydrodynamic equations the em potentials do not appear but only in local way through the strength of the em fields. In this way, the hydrodynamic equations exhibit more clearly the generation of the non-local character of quantum behavior than in the Schrödinger equation.
Even if the hydrodynamic and the wave descriptions are perfectly equivalent, no one prefers to solve the non-linear HQEs [1-3] instead of the Schrödinger one.
The mathematically more clear statements of non-local restrictions of the HQEs and their classical-like structure make the HQEs suitable for the achievement of the connection between quantum concepts (probabilities) and classical ones (e.g., trajectories) [4-6]. This fact makes the HQEs very useful in describing both phenomena at the edge between the quantum and classical mechanics such as the description of dispersive effects [7] critical phenomena [8], and other complex systems [9-10].
The advantage of HQEs in managing the non-local quantum character becomes more evident in system larger than a single atom when fluctuations becomes important [11] or when we want to investigate the effect of noise on the coherence of quantum non-local evolution [12], a field of great interest in the scientific community [13-18].
Since the non-local behavior of quantum mechanics (e.g., the superposition of states) is generated by the quantum potential, its relativistic expression coming from the quantum hydrodynamic description of the Dirac equations (DE) can be very useful in investigating the compatibility between the quantum non-local interactions and the relativistic postulate of finite speed of transmission of light and information. The Lorentz invariance of the quantum potential can give an important contribution to the solution of the problem of superluminal transmission of information in quantum mechanics [19-20] that has been postulated in order to overcome the contrast between the quantum phenomena and our sense of macroscopic reality [21-23].
The paper is organized as follows: in section 2 the hydrodynamic representation of the Dirac equation is derived; in section 3 it is calculated its low velocity limit and shown to agree with the hydrodynamic form of the Pauli equation; in section 4 the non-local property of quantum potential is discussed as well as its invariance under Lorentz transformation.

## 2. The hydrodynamic representation of the Dirac equation

Following the method used in a preceding paper [24], we proceed to find the current density conservation equation and the hydrodynamic force equation in agreement with the DE.
In relativistic mechanics it is well known that the DE

$$\left(i\hbar ¥^{«} \partial_{«} + mc\right)\Psi = 0 \tag{1}$$

by the minimal coupling with the electromagnetic field) reads

$$\left(i\hbar ¥^{«}\left(\partial_{«} + \frac{ie}{\hbar}A_{«}\right) + mc\right)\Psi = 0 \tag{2}$$

where « is a four-dimensional index that for the space-time vector reads $q_{«} = (ct, iq_j)$, where $q_j$ are the spatial components, and where

$$\partial_{«} = \left(\frac{1}{c}\frac{\partial}{\partial t}, i\nabla\right),\quad \Psi = \begin{pmatrix} \Psi_1 \\ \Psi_2 \\ \Psi_3 \\ \Psi_4 \end{pmatrix},\quad \left(¥^{«}\right) = \begin{pmatrix} 0 & \tilde{±}^{«} \\ ±^{«} & 0 \end{pmatrix},$$

where

$$\left(±^{0}\right) = \begin{pmatrix} 1 & 0 \\ 0 & 1 \end{pmatrix} = \left(\tilde{±}^{0}\right) \tag{3.a}$$

$$\left(±^{1}\right) = \begin{pmatrix} 0 & 1 \\ 1 & 0 \end{pmatrix} = -\left(\tilde{±}^{1}\right) \tag{3.b}$$

$$\left(±^{2}\right) = \begin{pmatrix} 0 & -i \\ i & 0 \end{pmatrix} = -\left(\tilde{±}^{2}\right) \tag{3.c}$$

$$\left(±^{3}\right) = \begin{pmatrix} 1 & 0 \\ 0 & -1 \end{pmatrix} = -\left(\tilde{±}^{3}\right). \tag{3.d}$$

Moreover, (2) can be re-cast in the Schrödinger-like form

$$i\hbar\partial_t\Psi = H_D\Psi \tag{4}$$

where

$$H_D = c¥^{0}¥^{i} \bullet \left(\frac{\hbar}{i}\nabla - eA\right) + ¥^{0}mc^2 + e¤ \tag{5}$$

that for free particle reduces to

$$H_D = ¥^0 \left( ¥^i \bullet \frac{\hbar c}{i} \nabla + mc^2 \right) \qquad (6)$$

Moreover, given the property of the $\pm^{«}$ matrices, so that it holds both

$$\left(¥^{«}\right)^{\dagger} \left(\partial_{«}\right)^{\dagger} = ¥^{«} \partial_{«} , \qquad (7)$$

and

$$\partial_{«} \Psi^* ¥^{«} ¥^0 = \partial_{«} \Psi^* ¥^0 ¥^{«} = ¥^{«} \partial_{«} \Psi^* ¥^0 \qquad (8)$$

it follows that

$$\left( i\hbar ¥^{«} \left( \partial_{«} - \frac{ie}{\hbar} A_{«} \right) - mc \right) \overline{\Psi} = 0 \qquad (9)$$

that leads to the Schrödinger-like form

$$- i\hbar \partial_t \overline{\Psi} = H_D{}^* \overline{\Psi} \qquad (10)$$

where $\overline{\Psi} = \Psi^* ¥^0$ and where $H_D{}^* = c¥^0 ¥^i \bullet \left( -\frac{\hbar}{i} \nabla - eA \right) + ¥^0 mc^2 + e¤$ . In force of relations (2, 9) the current

$$J^{«} = \overline{\Psi} ¥^{«} \Psi \qquad (11)$$

obeys to the conservation equation

$$\begin{aligned} \partial_{«} J^{«} &= \partial_{«} \left( \overline{\Psi} ¥^{«} \Psi \right) = \overline{\Psi} \partial_{«} \left( ¥^{«} \Psi \right) + \partial_{«} \left( \overline{\Psi} ¥^{«} \right) \Psi \\ &= \frac{mc}{i\hbar} \overline{\Psi} \Psi - \overline{\Psi} ¥^{«} \frac{e}{i\hbar} A_{«} \Psi - \frac{mc}{i\hbar} \overline{\Psi} \Psi + \overline{\Psi} ¥^{«} \frac{e}{i\hbar} A_{«} \Psi = 0 \end{aligned} \qquad (12)$$

that by using the identities

$$c° = J^0 = \Psi^* ¥^0 ¥^0 \Psi = \Psi^* \Psi = \sum_{«=1}^{4} | \Psi_{«} |^2 = | \Psi |^2 \qquad (13)$$

$$J_i = \overline{\Psi} ¥^i \Psi = \Psi^* ¥^0 ¥^i \Psi \qquad (14)$$

leads to the conservation equation

$$\begin{aligned} \partial_{«} J^{«} &= \frac{1}{c} \frac{\partial | \Psi |^2}{\partial t} + \nabla \bullet J \\ &= \frac{\partial °}{\partial t} + \nabla \bullet ( ° \dot{q} ) = 0 \end{aligned} \qquad (15)$$

where

$$\dot{q}_i = \frac{\overline{\Psi} c \gamma^{i} \Psi}{|\Psi|^2} = \frac{\Psi^* c \gamma^{0} \gamma^{i} \Psi}{|\Psi|^2} \ . \tag{16}$$

It is useful for the calculations below to observe that, with the help of (4, 10), equation (12) leads to

$$\begin{aligned} & i\hbar\, \partial_t \left( \overline{\Psi} \gamma^{0} \Psi \right) \\ & = i\hbar \left( \overline{\Psi} \left( \partial_t \gamma^{0} \Psi \right) + \left( \partial_t \overline{\Psi} \gamma^{0} \right) \Psi \right) \\ & = \overline{\Psi} \gamma^{0} H_D \Psi - \left( H_D^{\,*} \overline{\Psi} \gamma^{0} \right) \Psi \end{aligned} \tag{17}$$

In order to end with the other independent hydrodynamic equations (to obtain the full quantum hydrodynamic representation as a function of $|\cdot|$ and $S$), we write the four-dimensional wave function as

$$\begin{aligned} \Psi & = \left( R_1\, exp[\tfrac{i}{\hbar} S_1],\quad R_2\, exp[\tfrac{i}{\hbar} S_2],\quad R_3\, exp[\tfrac{i}{\hbar} S_3],\quad R_4\, exp[\tfrac{i}{\hbar} S_4] \right) \\ & = \left[ R_1,\ R_2,\ R_3,\ R_4 \right] \begin{pmatrix} exp[\tfrac{i}{\hbar} S_1] & & & \\ & exp[\tfrac{i}{\hbar} S_2] & & \\ & & exp[\tfrac{i}{\hbar} S_3] & \\ & & & exp[\tfrac{i}{\hbar} S_4] \end{pmatrix} \\ & = \boldsymbol{R}\, exp[\tfrac{i}{\hbar} \boldsymbol{S}] \end{aligned} \tag{18}$$

where $R_i = |\Psi_i|$ are the components of the vector $\boldsymbol{R}$ and the matrix $\boldsymbol{S}$ reads

$$\boldsymbol{S} = \begin{pmatrix} S_1 & & & \\ & S_2 & & \\ & & S_3 & \\ & & & S_4 \end{pmatrix} . \tag{19}$$

Moreover, by multiplying (4, 8) by the matrices

$$\left[ \frac{1}{\Psi} \right] = \begin{bmatrix} \frac{1}{\Psi_1} & 0 & 0 & 0 \\ 0 & \frac{1}{\Psi_2} & 0 & 0 \\ 0 & 0 & \frac{1}{\Psi_3} & 0 \\ 0 & 0 & 0 & \frac{1}{\Psi_4} \end{bmatrix} \qquad \left[ \frac{1}{\overline{\Psi}} \right] = \begin{bmatrix} \frac{1}{\Psi_3 *} & 0 & 0 & 0 \\ 0 & \frac{1}{\Psi_4 *} & 0 & 0 \\ 0 & 0 & \frac{1}{\Psi_1 *} & 0 \\ 0 & 0 & 0 & \frac{1}{\Psi_2 *} \end{bmatrix} \tag{20}$$

we obtain the equation

$$\left( \left[ \frac{1}{\Psi} \right] \frac{\partial}{\partial t} \Psi - \left[ \frac{1}{\overline{\Psi}} \right] \frac{\partial}{\partial t} \overline{\Psi} \right) = \left( \frac{\partial}{\partial t} ln[\Psi] - \frac{\partial}{\partial t} ln[\overline{\Psi}] \right) \tag{21}$$

where $\frac{\partial}{\partial t} ln[\Psi] = \left( \frac{\partial}{\partial t} ln[\Psi_1], \quad \frac{\partial}{\partial t} ln[\Psi_2], \quad \frac{\partial}{\partial t} ln[\Psi_3], \quad \frac{\partial}{\partial t} ln[\Psi_4] \right)$ , that leads to (see appendix A)

$$\left( \frac{\partial}{\partial t} ln[\Psi] - \frac{\partial}{\partial t} ln[\overline{\Psi}] \right) = \frac{\partial}{\partial t} ln[\frac{\boldsymbol{R}}{\boldsymbol{R}\text{¥}^{0}}] + \frac{i}{\hbar} \left( \frac{\partial \boldsymbol{I} \left( \boldsymbol{S} + \boldsymbol{S}\text{¥}^{0} \right)}{\partial t} \right) \tag{22}$$
$$= -\frac{i}{\hbar} \left( \left[ \frac{1}{\Psi} \right] H_D \Psi + \left[ \frac{1}{\overline{\overline{\Psi}}} \right] H_D{}^{*} \overline{\Psi} \right)$$

where $\frac{1}{\overline{\overline{\Psi}}} = \frac{1}{\Psi * \text{¥}^{0}}$ , $\boldsymbol{I} = (1,1,1,1)$ and where

$$\frac{\partial}{\partial t} ln[\frac{\boldsymbol{R}}{\boldsymbol{R}\text{¥}^{0}}] = \left( \frac{\partial}{\partial t} ln[\frac{R_1}{R_3}], \quad \frac{\partial}{\partial t} ln[\frac{R_2}{R_4}], \quad \frac{\partial}{\partial t} ln[\frac{R_3}{R_1}], \quad \frac{\partial}{\partial t} ln[\frac{R_4}{R_2}] \right) \tag{23}$$

and, hence, to (see appendix B)

$$\frac{\hbar}{i} \frac{\partial}{\partial t} ln[\frac{\boldsymbol{R}}{\overline{\overline{\boldsymbol{R}}}}] + \left( \frac{\partial \boldsymbol{I} \left( \boldsymbol{S} + \overline{\boldsymbol{S}} \right)}{\partial t} \right) = -c\text{¥}^{0}\text{¥}^{i} \bullet \left( \nabla \left( \boldsymbol{S} + \overline{\boldsymbol{S}} \right) - 2eA \right) \tag{24}$$
$$- c\text{¥}^{0}\text{¥}^{i} \bullet \frac{\hbar}{i} \left( \nabla\, ln[\boldsymbol{R}] - \nabla\, ln[\overline{\boldsymbol{R}}] \right) - 2\left( \text{¥}^{0} mc^{2} + e\text{¤} \right) .$$

where $\overline{\boldsymbol{R}} = \boldsymbol{R}\text{¥}^{0}$, $\overline{\boldsymbol{S}} = \boldsymbol{S}\text{¥}^{0}$,

$$\left[ \frac{1}{\boldsymbol{R}} \right] = \begin{bmatrix} \frac{1}{\boldsymbol{R}_1} & 0 & 0 & 0 \\ 0 & \frac{1}{\boldsymbol{R}_2} & 0 & 0 \\ 0 & 0 & \frac{1}{\boldsymbol{R}_3} & 0 \\ 0 & 0 & 0 & \frac{1}{\boldsymbol{R}_4} \end{bmatrix} \tag{25.a}$$

and where

$$\left[ \frac{1}{\overline{\overline{\boldsymbol{R}}}} \right] = \begin{bmatrix} \frac{1}{\boldsymbol{R}_3} & 0 & 0 & 0 \\ 0 & \frac{1}{\boldsymbol{R}_4} & 0 & 0 \\ 0 & 0 & \frac{1}{\boldsymbol{R}_1} & 0 \\ 0 & 0 & 0 & \frac{1}{\boldsymbol{R}_2} \end{bmatrix} .. \tag{25.b}$$

Moreover, by multiplying (24) on the left by $\frac{\Psi *}{|\Psi|}$ and on the right by $\frac{\Psi}{|\Psi|}$, it follows that

$$\frac{\partial \boldsymbol{I}\left(\boldsymbol{S}+\overline{\boldsymbol{S}}\right)}{\partial t}\frac{\Psi * \Psi}{|\Psi|^2} = -\frac{\Psi * c ¥^0 ¥^i \Psi}{|\Psi|^2} \bullet \left(\nabla \boldsymbol{I}\left(\boldsymbol{S}+\overline{\boldsymbol{S}}\right) - 2eA\right) - 2\frac{\Psi * \Psi}{|\Psi|^2}\left(¥^0 mc^2 + e¤\right)$$
$$-\frac{\Psi * c ¥^0 ¥^i \Psi}{|\Psi|^2} \bullet \frac{\hbar}{i}\nabla\, ln[\frac{\boldsymbol{R}}{\overline{\overline{\boldsymbol{R}}}}] - \frac{\Psi * \Psi}{|\Psi|^2}\frac{\hbar}{i}\frac{\partial}{\partial t} ln[\frac{\boldsymbol{R}}{\overline{\overline{\boldsymbol{R}}}}] \tag{26}$$

where

$$\nabla\, ln[\frac{\boldsymbol{R}}{\overline{\overline{\boldsymbol{R}}}}] = \left(\nabla\, ln[\frac{R_1}{R_3}],\quad \nabla\, ln[\frac{R_2}{R_4}],\quad \nabla\, ln[\frac{R_3}{R_1}],\quad \nabla\, ln[\frac{R_4}{R_2}]\right), \tag{27}$$

that since $\frac{\Psi * \Psi}{|\Psi|^2} = 1$ (by (16)), leads to

$$\frac{\partial \boldsymbol{I}\left(\boldsymbol{S}+\overline{\boldsymbol{S}}\right)}{\partial t} + 2\left(¥^0 mc^2 + e¤\right) + \dot{q}\bullet \nabla \boldsymbol{I}\left(\boldsymbol{S}+\overline{\boldsymbol{S}}\right) - \dot{q}\bullet 2eA = -\frac{\hbar}{i}\left[\dot{q}\bullet \nabla\, ln[\frac{\boldsymbol{R}}{\overline{\overline{\boldsymbol{R}}}}] + \frac{\partial}{\partial t} ln[\frac{\boldsymbol{R}}{\overline{\overline{\boldsymbol{R}}}}]\right] \tag{28}$$

Moreover, by taking the gradient of equation (28), it follows that

$$\frac{\partial\left(\nabla \boldsymbol{I}\left(\boldsymbol{S}+\overline{\boldsymbol{S}}\right) - 2eA\right)}{\partial t} + 2e\left(\nabla ¤ + \frac{\partial A}{\partial t}\right) + 2¥^0 c^2 \nabla m - \nabla \dot{q}\bullet \nabla \boldsymbol{I}\left(\boldsymbol{S}+\overline{\boldsymbol{S}}\right)$$
$$-\dot{q}\bullet \nabla\left(\nabla \boldsymbol{I}\left(\boldsymbol{S}+\overline{\boldsymbol{S}}\right) - 2eA\right) - 2e\dot{q}\bullet \nabla A + 2e\dot{q}_i \nabla A_i + \nabla \dot{q}_i\left(\partial_i \boldsymbol{I}\left(\boldsymbol{S}+\overline{\boldsymbol{S}}\right) - eA_i\right) \tag{29}$$
$$= \frac{\hbar}{i}\nabla\left[\dot{q}\bullet\left(\frac{1}{\boldsymbol{R}}\nabla \boldsymbol{R} - \frac{1}{\overline{\overline{\boldsymbol{R}}}}\nabla \overline{\boldsymbol{R}}\right) + \frac{\partial}{\partial t} ln[\frac{\boldsymbol{R}}{\overline{\overline{\boldsymbol{R}}}}]\right]$$

where it has been used the property that the rotor of the gradient of the action is null (i.e., the quantization condition) [2]

$$\nabla \times \nabla \boldsymbol{I}\left(\boldsymbol{S}+\overline{\boldsymbol{S}}\right) = 0\,. \tag{30}$$

Furthermore, by posing

$$\boldsymbol{p} = \frac{\nabla \boldsymbol{I}\left(\boldsymbol{S}+\overline{\boldsymbol{S}}\right)}{2} \tag{31}$$

equation (29) reads (see appendix C)

$$\frac{d\left(\boldsymbol{p} - eA\right)}{dt} = -\mathrm{e}\left(\boldsymbol{E} + \dot{q}\times \boldsymbol{B}\right) - ¥^0 c^2 \nabla m - \nabla \dot{q}_i\left(\boldsymbol{p} - eA\right)_i$$
$$-\frac{\hbar}{2i}\nabla\left[\dot{q}\bullet \nabla\, ln[\frac{\boldsymbol{R}}{\overline{\overline{\boldsymbol{R}}}}] + \frac{\partial}{\partial t} ln[\frac{\boldsymbol{R}}{\overline{\overline{\boldsymbol{R}}}}]\right] \tag{32}$$

that using the identity

$$\dot{q}_{«} = \frac{\overline{\Psi} c ¥^{«} \Psi}{|\Psi|^2} = \frac{\Psi * c ¥^0 ¥^{«} \Psi}{|\Psi|^2} \tag{33}$$

finally leads to

$$\frac{d(\boldsymbol{p}-eA)}{dt} = -\mathsf{e}\left(\boldsymbol{E}+\dot{q}\times\boldsymbol{B}\right) - ¥^{0}c^{2}\nabla m - \nabla\dot{q}_{i}(\boldsymbol{p}-eA)_{i}$$

$$+\frac{\hbar}{2i}\nabla\left[\dot{q}^{«}\ \partial_{«}\ ln[\frac{\boldsymbol{R}}{\boldsymbol{R}}]\right] \tag{34}$$

If, by using the correspondence rules $\boldsymbol{p} \rightarrow -i\hbar\nabla$ , we derive the Dirac Hamiltonian as a function of the hydrodynamic variables $(\boldsymbol{q},\boldsymbol{p})$

$$H_{D} = c¥^{0}¥^{i}\bullet(\boldsymbol{p}-eA)+¥^{0}mc^{2}+e¤ \tag{35.a}$$

and its density

$$\mathcal{H}_{D} = \frac{\Psi^{*}H_{D}\Psi}{|\Psi|^{2}} = \dot{q}\bullet(\boldsymbol{p}-eA)+¥^{0}mc^{2}+e¤ \tag{35.b}$$

where $\dot{q}$ is given by (16), we can write the hydrodynamic Hamiltonian-like relations that read

$$\dot{q}_{H} = \frac{\partial\mathcal{H}_{D}}{\partial\boldsymbol{p}} = \dot{q} \tag{36}$$

leading to autonomous velocity since $\dot{q} = \dot{q}_{(q,t)}$ is decoupled by the moment, and

$$-\nabla(\mathcal{H}_{D}+V_{qu}) = \dot{p} = -\nabla\left(\dot{q}\bullet(p-eA)+¥^{0}mc^{2}+e¤+V_{qu}\right) \quad . \tag{37}$$

From (37) the total derivative of the kinetic moment reads

$$\frac{d(\boldsymbol{p}-eA)}{dt} = -\left(\mathsf{e}\nabla¤+¥^{0}c^{2}\nabla m+\mathsf{e}\left(\dot{q}_{i}\ \nabla A_{i}-\frac{\partial A}{\partial t}-\dot{q}\bullet\nabla A\right)+\nabla\dot{q}_{i}(\boldsymbol{p}-eA)_{i}\right)-\nabla V_{qu}$$

$$= -\mathsf{e}\left(\boldsymbol{E}+\dot{q}\times\nabla\times A\right)-¥^{0}c^{2}\nabla m-\nabla\dot{q}_{i}(\boldsymbol{p}-eA)_{i}-\nabla V_{qu} \tag{38}$$

$$= -\mathsf{e}\left(\boldsymbol{E}+\dot{q}\times\boldsymbol{B}\right)-¥^{0}c^{2}\nabla m-\nabla\dot{q}_{i}(\boldsymbol{p}-eA)_{i}-\nabla V_{qu}$$

from which we obtain equation (34) through the expression of the relativistic quantum potential $V_{qu}$ (RQP) that reads

$$V_{qu} = -\frac{\hbar}{2i}\left[\dot{q}^{«}\ \partial_{«}\ ln[\frac{\boldsymbol{R}}{\boldsymbol{R}}]\right] \quad . \tag{39}$$

By setting to zero the RQP it follows that the motion of the density ° is defined by a local equation of motion describing the evolution of a relativistic classical dust.

Moreover, in order to investigate the dynamics of such a local dust °, it is possible to define the corresponding non-linear Dirac equation $i\hbar\partial_t\Psi = H_{D_{cl}}\Psi$ (similarly to the nonlinear Schrödinger equation [25]) by subtracting the non-local quantum potential, to obtain

$$H_{D_{cl}} = c¥^0¥^i \bullet \left(\frac{\hbar}{i}\nabla - eA\right) + ¥^0 mc^2 + e¤ + i\hbar \frac{\overline{\Psi}}{2|\Psi|^2}\left(\dot{q}^{«}\ \partial_{«}\ ln[|\frac{\Psi}{\overline{\Psi}}|]\right)\Psi \qquad (40.a)$$

$$H_{D_{cl}} = c¥^0¥^i \bullet \left(\frac{\hbar}{i}\nabla - eA\right) + ¥^0 mc^2 + e¤ + \frac{i\hbar}{2}\dot{q}^{«}\ \partial_{«}\ ln[|\frac{\Psi}{\overline{\Psi}}|] \qquad (40.b)$$

$$H_{D_{cl}} = c¥^0¥^i \bullet \left(\frac{\hbar}{i}\nabla - eA\right) + ¥^0 mc^2 + e¤ + \frac{i\hbar c}{2}\frac{\Psi * ¥^0¥^{«}\Psi}{|\Psi|^2}\partial_{«}\ ln[|\frac{\Psi}{\overline{\Psi}}|] \qquad (40.c)$$

where $ln[|\frac{\Psi}{\overline{\Psi}}|] = \left( ln[\frac{|\Psi_1|}{|\Psi_3*|}],\ ln[\frac{|\Psi_2|}{|\Psi_4*|}]\ \ ln[\frac{|\Psi_3|}{|\Psi_1*|}]\ \ ln[\frac{|\Psi_4|}{|\Psi_2*|}] \right)$.

Furthermore, if we consider the more realistic case where noise is present (e.g., we may consider the sufficiently general case of the Gaussian one) we have the stochastic equation [12]

$$\frac{\partial °}{\partial t} + \nabla \bullet (°\,\dot{q}) + ¦_{(q,t,T)} = 0 \qquad (41)$$

where the Gaussian noise is defined by its variance that in a sufficiently general form can read

$$\lim_{T\to 0} <¦_{(q_{~},t)}, ¦_{(q\ +ª,t+²)}> = <¦_{(q_{~},t)}, ¦_{(q_{~},t)}>_{(T)}\ G(\frac{ª}{ª_{c(T)}})¢(²)¢_{~} \qquad (42)$$

where T is the amplitude of noise (e.g., the temperature of the ideal gas thermostat). [12] and $ª_c$ is the correlation length of the Gaussian noise. If we translate (41) back to the Dirac formalism we obtain

$$\left( i\hbar¥^{«}\left(\partial_{«} + \frac{ie}{\hbar}A_{«}\right) + mc + \frac{i}{c}\frac{¥^0}{|\Psi|^2}¦_{(q,t,T)} \right)\Psi = 0 \qquad (43)$$

the represents the stochastic analog of the Dirac equation that in the Schrödinger-like form $i\hbar\partial_t\Psi = H_D\Psi$ leads to

$$H_D = c¥^0¥^i \bullet \left(\frac{\hbar}{i}\nabla - eA\right) + ¥^0 mc^2 + e¤ + \frac{i}{|\Psi|^2}¦_{(q,t,T)} \qquad (44)$$

In principle, both the correlation length $ª_c$ and the correlation function $G(\frac{ª}{ª_c})$ of the Gaussian noise are free parameters but if we add the additional constraint that the (mean square root of) energy fluctuations of the quantum potential must remain finite (needed in the stochastic case to exclude non-physical solutions [12]), a condition comes on them. In the classical limit it has been shown [12] that, in the small noise amplitude limit, the correlation length $a_c$ as well as $G(\frac{a}{a_c})$ acquire the expressions

$$\lim_{T \to 0} G(\frac{a}{a_c}) = exp[-(\frac{a}{a_c})^2 ] \tag{45}$$

with

$$\lim_{T \to 0} a_c = (\frac{®}{2})^{3/2} \frac{\hbar}{(2mkT)^{1/2}} \tag{46}$$

The additional condition of non-diverging energy is needed in the stochastic case since the quantum potential is critically dependent by the distance on which independent fluctuations happen. Fluctuation of density ° with null correlation distance brings to infinite quantum potential energy. This is due to the derivative form of the quantum potential whose energy is given by the partial derivative of the wave function modulus.

# 3. The classical limit

In order to derive the classical limit, we use the following limiting expression for $H_D$ [2] that reads

$$H_D \cong \begin{bmatrix} H_{D+} & 0 \\ 0 & H_{D_-} \end{bmatrix} \tag{47}$$

where

$$\begin{aligned} H_{D\pm} &\cong \pm mc^2 + e¤ + \frac{(\boldsymbol{\sigma} \bullet ®)^2}{2m} \\ &\cong \pm m_0 c^2 \pm \frac{m_0 \dot{q}^2}{2} + e¤ + \frac{e\hbar \boldsymbol{\sigma} \bullet B}{2m} \\ &\cong \pm m_0 c^2 \pm \frac{®^2}{2m_0} + e¤ + «\boldsymbol{\sigma} \bullet B \\ &= \pm m_0 c^2 \pm \frac{1}{2m}(\frac{\hbar}{i}\nabla - eA)^2 + e¤ + «\boldsymbol{\sigma} \bullet B \end{aligned} \tag{48}$$

where $\boldsymbol{\sigma} = (\pm_1, \pm_2, \pm_3)$ and $® = (\boldsymbol{p} - eA)$. Moreover, being $H_{D\pm}$ real, so that it holds

$$-i\hbar\partial_t \Psi^* = H_D{}^* \Psi^*, \tag{49}$$

the current conservation equation (12) reads

$$i\hbar(\partial_t \Psi^* \Psi) = i\hbar c(\partial_t °) = \Psi^* H_D \Psi - (H_D{}^* \Psi^*)\Psi \tag{50}$$

That by the diagonal form of $H_D$ (since particle and antiparticle are decoupled) and by using the notation

$$\Psi = (\Psi_+, \Psi_-) \text{ where} \left(\Psi_\pm\right) = |\Psi_\pm| \, exp[\frac{iS}{\hbar}] = ¡_\pm \, |\cdot| \, exp[\frac{iS}{\hbar}] \,, \tag{51}$$

where $¡_\pm$ it is a bi-dimensional spinor, leads to the expression

$$i\hbar \frac{\partial}{\partial t} |\Psi_\pm|^2 = \frac{1}{2m} \left( \Psi^\dagger{}_\pm (\frac{\hbar}{i} \nabla - eA)^2 \Psi_\pm - \Psi_\pm (\frac{\hbar}{i} \nabla - eA)^2 \Psi^\dagger{}_\pm \right)$$
$$= -\frac{\hbar^2}{2m} \nabla \left( \Psi^\dagger{}_\pm \nabla \Psi_\pm - \left( \nabla \Psi^\dagger \right) \Psi_\pm \right) - 2 \frac{\hbar}{2im} \nabla \bullet |\Psi_\pm|^2 \, eA \quad , \tag{52}$$

where being $\Psi_-$ and $\Psi_+$ decoupled, we can consider just that one with the plus sign to obtain (see appendix D)

$$\frac{\partial}{\partial t} |\cdot|^2 = -\nabla \bullet \left( |\cdot|^2 \frac{1}{m} \nabla S \right) + \frac{1}{m} \nabla \bullet |\cdot|^2 \, eA + \frac{\hbar}{2m} \nabla \bullet \left( |\cdot|^2 \, cos\, \text{g} \nabla ¨ \right)$$
$$= -\nabla \bullet \left( |\cdot|^2 \left( \frac{(\nabla S - eA)}{m} + \frac{\hbar}{2m} cos\, \text{g} \nabla ¨ \right) \right) \tag{53}$$

where it has been used the notation

$$\left(\Psi\right) = \left(¡\right) |\cdot| \, exp[\frac{iS}{\hbar}] \,, \tag{54}$$

and where

$$¡ = \begin{pmatrix} cos\left(\frac{\text{g}}{2}\right) exp[-i \frac{¨}{2}] \\ sin\left(\frac{\text{g}}{2}\right) exp[i \frac{¨}{2}] \end{pmatrix} \tag{55}$$

where g and ¨ are the angles in spherical co-ordinates of the versor $\boldsymbol{n} = \frac{\cdot^\dagger \boldsymbol{\sigma} \cdot}{\cdot^\dagger \cdot}$ (defining the magnetic moment density $« |\cdot|^2 \boldsymbol{n}$ ).

Moreover, since in the classical limit particles and antiparticles are decoupled, we can factorize the mass phase factor as

$$\left(\Psi\right)_\pm = exp[\pm \left( -\frac{imc^2 t}{\hbar} \right)] \left(\Psi_{cl}\right)_\pm \tag{56}$$

with the re-defined Hamiltonian

$$H_{D\pm} \cong \pm \frac{1}{2m} (\frac{\hbar}{i} \nabla - eA)^2 + e¤ + « \boldsymbol{\sigma} \bullet B \,. \tag{57}$$

and the re-defined action

$$S = S_{cl} - mc^2 t \,. \tag{58}$$

Thence, in the low speed limit, the conservation equation reads

$$\frac{\partial}{\partial t}|\cdot\;|^2 = -\nabla \bullet \left( |\cdot\;|^2 \overset{\bullet}{q} \right). \qquad (59)$$

with

$$\overset{\bullet}{q} = \frac{1}{m}\left( (\nabla S_{cl} - eA) + \frac{\hbar}{2} cos\, \mathrm{g} \nabla \ddot{\ } \right) \qquad (60)$$

where it has been introduced the approximation $\nabla mc^2 t \cong 0$ , since in the low velocity limit the particle mass can be considered constant, and where it has been used the property $|\Psi| = |\cdot\;|$ (being $|\text{¡}|^2 = 1$).
Equation (59) agrees with the wave function density conservation equation of the hydrodynamic representation of the Pauli's equation [2].
The hydrodynamic-like force equation acting on the particle density and spin can be obtained by multiplying on the left side the Dirac equations (4, 10) by the matrices $\left[\frac{1}{\Psi}\right]$ and $\left[\frac{1}{\overline{\overline{\Psi}}}\right]$, respectively, leading to the following relation

$$\left( \left[\frac{1}{\Psi}\right] \frac{\partial}{\partial t}\Psi - \left[\frac{1}{\overline{\overline{\Psi}}}\right] \frac{\partial}{\partial t}\overline{\Psi} \right) = \left( \frac{\partial}{\partial t} ln[\Psi] - \frac{\partial}{\partial t} ln[\overline{\Psi}] \right)$$
$$= \frac{\partial}{\partial t} ln[\frac{\Psi}{\overline{\overline{\Psi}}}] = -\frac{i}{\hbar}\left( \left[\frac{1}{\Psi}\right] H_D \Psi + \left[\frac{1}{\overline{\overline{\Psi}}}\right] H_D{}^{*} \overline{\Psi} \right) \qquad (61)$$

where in the low velocity limit it follows that $\overline{\Psi} \to \Psi^*$ , where

$$= \begin{pmatrix} I & \\ & -I \end{pmatrix}, \qquad (62)$$

where $I = \begin{pmatrix} 1 & 0 \\ 0 & 1 \end{pmatrix}$ and where

$$H_D \cong \frac{1}{2m}(\frac{\hbar}{i}\nabla - eA)^2 + m_0 c^2 + e\text{¤} + \text{«}\boldsymbol{\sigma} \bullet B, \qquad (63)$$

$$\left[\frac{1}{\Psi}\right]=\begin{bmatrix}\frac{1}{\Psi_1} & 0 & 0 & 0\\ 0 & \frac{1}{\Psi_2} & 0 & 0\\ 0 & 0 & \frac{1}{\Psi_3} & 0\\ 0 & 0 & 0 & \frac{1}{\Psi_4}\end{bmatrix}\rightarrow\begin{bmatrix}\frac{1}{\Psi_{1+}} & 0 & 0 & 0\\ 0 & \frac{1}{\Psi_{2_+}} & 0 & 0\\ 0 & 0 & \frac{1}{\Psi_{1-}} & 0\\ 0 & 0 & 0 & \frac{1}{\Psi_{2_-}}\end{bmatrix}$$

$$=\begin{bmatrix}\left[\frac{1}{\Psi_+}\right] & \begin{matrix}0 & 0\\ 0 & 0\end{matrix}\\ \begin{matrix}0 & 0\\ 0 & 0\end{matrix} & \left[\frac{1}{\Psi_-}\right]\end{bmatrix} \qquad (64)$$

By considering just the positive-energy spinor and eliminating the mass phase factor by using the Hamiltonian (57), relation (61) reads

$$\frac{\partial}{\partial t}ln[\frac{\Psi}{\Psi^*}]=-\frac{i}{\hbar}\left(\left[\frac{1}{\Psi}\right]H_{D_+}\Psi+\left[\frac{1}{\Psi^*}\right]H_{D_+}{}^*\Psi^*\right)$$

$$=-\frac{i}{\hbar}\left(\begin{matrix}2e¤+\frac{(eA)^2}{m}-\frac{\hbar^2}{2m}\left(\left[\frac{1}{\Psi}\right]\nabla\bullet\nabla\Psi-\left[\frac{1}{\Psi^*}\right]\nabla\bullet\nabla\Psi^*\right)\\ -\frac{\hbar}{2im}\left(\left[\frac{1}{\Psi}\right]e(\nabla\bullet A\Psi)-\left[\frac{1}{\Psi^*}\right]e(\nabla\bullet A\Psi *)\right)\\ -\frac{\hbar}{2im}\left(\left[\frac{1}{\Psi}\right]eA\bullet\nabla\Psi-\left[\frac{1}{\Psi^*}\right]eA\bullet\nabla\Psi^*\right)\end{matrix}\right) \qquad (65)$$

$$-\frac{i}{\hbar}\left(\left[\frac{1}{\Psi}\right](«\boldsymbol{\sigma}\bullet B)\Psi+\left[\frac{1}{\Psi^*}\right](«\boldsymbol{\sigma}\bullet B)\Psi^*\right)$$

where

$$\frac{\partial}{\partial t}ln[\frac{\Psi}{\Psi^*}]=\left(\frac{\partial}{\partial t}ln[\frac{\Psi_1}{\Psi_1 *}],\ \ \frac{\partial}{\partial t}ln[\frac{\Psi_2}{\Psi_2 *}]\right)$$

$$=\left(\frac{\partial}{\partial t}\left(ln[\frac{¡_1}{¡_1 *}]+2i\frac{S}{\hbar}\right),\ \ \frac{\partial}{\partial t}\left(ln[\frac{¡_2}{¡_2 *}]+2i\frac{S}{\hbar}\right)\right)$$

$$=2\frac{i}{\hbar}\left(\frac{\partial}{\partial t}(-\hbar¨+S),\ \ \frac{\partial}{\partial t}\left(\hbar\frac{¨}{2}+S\right)\right) \qquad (66)$$

$$=2\frac{i}{\hbar}\left((1,1)\frac{\partial S}{\partial t}+(-1,1)\hbar\frac{\partial ¨}{\partial t}\right)$$

For sake of simplicity (see appendix E) we will give here the solution for spinless particles obtained for $« = 0$ and $¡ = constant$ (so that it holds $\nabla ¡ = 0$ and $\nabla ¨ = \nabla g = 0$ ) that leads to

$$\left((1,1)\frac{\partial S}{\partial t}+(-1,1)\frac{\partial\hbar\,\text{¨}}{\partial t}\right)$$

$$=-\frac{1}{2}(1,1)\left(\begin{array}{l} 2e\text{¤}+\frac{(eA)^2}{m}-\frac{\hbar^2}{2m}\left(\left[\frac{1}{\cdot}\right]\nabla\bullet\nabla\cdot\;\;-\left[\frac{1}{\cdot\,^*}\right]\nabla\bullet\nabla\cdot\;^*\right) \\ -\frac{\hbar}{2im}\left(\left[\frac{1}{\cdot}\right]e(\nabla\bullet A\cdot\;)-\left[\frac{1}{\cdot\,^*}\right]e(\nabla\bullet A\cdot\;*)\right) \\ -\frac{\hbar}{2im}\left(\left[\frac{1}{\cdot}\right]eA\bullet\nabla\cdot\;\;-\left[\frac{1}{\Psi^*}\right]eA\bullet\nabla\cdot\;^*\right) \end{array}\right) \tag{66}$$

$$=-\frac{1}{2}(1,1)\left(\begin{array}{l} 2e\text{¤}+\frac{(eA)^2}{m}-\frac{\hbar^2}{2m}|\cdot\;|^{-1}\left(\begin{array}{l} exp[-\frac{iS}{\hbar}]\nabla\bullet\nabla\left(|\cdot\;|exp[\frac{iS}{\hbar}\right) \\ -exp[+\frac{iS}{\hbar}]\nabla\bullet\nabla\left(|\cdot\;|exp[-\frac{iS}{\hbar}\right) \end{array}\right) \\ -\frac{\hbar}{2im}\left(\left[\frac{1}{\cdot}\right]e(\nabla\bullet A\cdot\;)-\left[\frac{1}{\cdot\,^*}\right]e(\nabla\bullet A\cdot\;*)\right) \\ -\frac{\hbar}{2im}eA\bullet\left(\left[\frac{1}{\cdot}\right]\nabla\cdot\;\;-\left[\frac{1}{\cdot\,^*}\right]\nabla\cdot\;^*\right) \end{array}\right) \tag{68}$$

$$=-\frac{1}{2}(1,1)\left(\begin{array}{l} 2e\text{¤}+\frac{(eA)^2}{m}-\frac{\hbar^2}{2m}|\cdot\;|^{-1}\left(\begin{array}{l} exp[-\frac{iS}{\hbar}]\nabla\bullet\nabla\left(|\cdot\;|exp[\frac{iS}{\hbar}\right) \\ -exp[+\frac{iS}{\hbar}]\nabla\bullet\nabla\left(|\cdot\;|exp[-\frac{iS}{\hbar}\right) \end{array}\right) \\ -2\frac{eA\bullet\nabla S}{m} \end{array}\right) \tag{69}$$

$$\left((1,1)\frac{\partial S}{\partial t}+(-1,1)\hbar\frac{\partial\,\text{¨}}{\partial t}\right)=-(1,1)\left(e\text{¤}+\frac{(\nabla S-eA)^2}{2m}-\frac{\hbar^2}{2m}|\cdot\;|^{-1}\nabla^2|\cdot\;|\right) \tag{70}$$

and to

$$\left((1,1)\frac{\partial\nabla S}{\partial t}+(-1,1)\hbar\frac{\partial\nabla\,\text{¨}}{\partial t}\right)$$

$$=-(1,1)\left(\begin{array}{l} e\nabla\text{¤}-\frac{\hbar^2}{2m}\nabla\left(|\cdot\;|^{-1}\nabla^2|\cdot\;|\right)+\frac{(\nabla S-eA)\bullet\nabla(\nabla S-eA)}{m} \\ +\left(\frac{(\nabla S-eA)_i\nabla(\nabla S-eA)_i}{m}-\frac{(\nabla S-eA)\bullet\nabla(\nabla S-eA)}{m}\right) \end{array}\right). \tag{71}$$

Moreover, by using the relation (60) with the condition $\nabla ¨ = 0$ (i.e., $\dot{q} = \frac{1}{m}\left(\nabla S - eA\right)$), relation (71) leads to

$$(1,1)\frac{\partial \nabla S}{\partial t} = -(1,1)\begin{pmatrix} e\nabla ¤ + \dot{q}\bullet\nabla\left(\nabla S - eA\right) - e\left(\dot{q}\times\nabla\times A\right) \\ +\left(\dot{q}\times\nabla\times\nabla S\right) - \frac{\hbar^{2}}{2m}\nabla\left(|\cdot|^{-1}\nabla^{2}|\cdot|\right)\end{pmatrix} \tag{72}$$

and, being the gradient of the action irrotational [2] (i.e., $\nabla\times\nabla S = 0$), to

$$\left(\frac{\partial \nabla S}{\partial t}\right) = -\left(e\nabla ¤ + \dot{q}\bullet\nabla\left(\nabla S - eA\right) - e\left(\dot{q}\times\nabla\times A\right) - \frac{\hbar^{2}}{2m}\nabla\left(|\cdot|^{-1}\nabla^{2}|\cdot|\right)\right) \tag{73}$$

that finally reads

$$\left(\frac{d\left(\nabla S - eA\right)}{dt}\right) = -\left(e\left(\nabla ¤ + \frac{\partial A}{\partial t}\right) - e\left(\dot{q}\times\boldsymbol{B}\right) - \frac{\hbar^{2}}{2m}\nabla\left(|\cdot|^{-1}\nabla^{2}|\cdot|\right)\right) \tag{74}$$

leading to the expression

$$\frac{d\dot{q}}{dt} = \frac{e}{m}\left(\boldsymbol{E} + \dot{q}\times\boldsymbol{B}\right) + \frac{\hbar^{2}}{2m^{2}}\nabla\left(|\cdot|^{-1}\nabla^{2}|\cdot|\right) \tag{75}$$

that represents the correct classical limit [2].

## 4. Discussion

If we look at the manageability of the quantum equations no one would solve the hydrodynamic ones. Nevertheless, the interest for the QHA remained unaltered since it was proposed by Madelung [3]. The motivation comes both from the formal analogy with the classical mechanics and from the fact that the QHA model facilitates the correlation between the quantum and the classical dynamics since the non local properties of quantum mechanics can be more easily recognizable in it.
If in the Schrödinger problem not all solutions are considered, but only those that fulfill precise boundary conditions (e.g., for bounded problems the eigenstates are those that go to zero to infinity) so that the quantization comes in, in the QHA, these non local characteristics are transferred to the dynamics through the quantum potential (39).
This is clearly recognizable in the classical limit where, if we subtract the contribution of the quantum potential to the quantum equation, the classic non-linear Schrödinger one is obtained [25].
On the other hand, if the quantum potential is null, the hydrodynamic equations describe the motion of a classical dust of density ° [2].
In the QHA the eigenstates are defined by the stationary densities that happen when the force generated by the quantum potential exactly counterbalances that one due to the Hamiltonian potential (with the initial condition $\dot{q} = 0$).
Since the quantum potential changes with the state of the system, more than one stationary state is possible.
If we disregard the quantum potential, we also wipe out the quantum eigenstates (with the consequent quantum superposition of states) and we end with the classical equation of motion.

Thence, it clearly comes out that in the QHA the non locality does not come from boundary conditions (that are apart from the equations) but from the quantum pseudo-potential that depends by the state of the system and is a source of an elastic-like (non-local) energy [12,25,26].
Just for instance, if we consider a bi-dimensional space, the quantum potential makes the vacuum acting like an elastic membrane that becomes quite rigid against curvature on very small scale.
Since the force of the quantum potential in a point depends by the state of the system around it, the character of non-local dynamics is introduced into the QHA equations.
The fact that the state of a quantum system in a point depends by its state in far away regions generates a rejection from our sense of reality.
The determination of the result of a quantum measurement as a function of what happen to a quantum entangled state at a far distance is at the base of the Einstein-Podolsky-Rosen paradox [21] where the classic concepts conflict with the quantum results.
Nevertheless, Bell' inequalities and connected works [22-23] compared with experiments show that the Copenhagen interpretation of quantum mechanics is always verified.
The fact that the result of a quantum measurement is determined by what happens far away (in another experiment) has led many physicists to postulate the possibility of quantum transmission of information at a speed larger than that one of the light [19-20].
The availability of the relativistic quantum potential (39) allows verifying if the non-local interactions involved in the quantum mechanics propagate themselves compatibly with the postulate of the relativity about the invariance of light speed as the fastest way to which signals and interactions are transmitted.
Since the invariance of light speed is the generating property of the Lorentz transformations, the invariance of quantum potential under the same transformation allows affirming that the quantum non-local behavior is compatible with such a postulate of the relativity.
If we analyze the expression (39)

$$V_{qu} = -\frac{\hbar}{2i}\left[ \dot{q}^{«} \;\; \partial_{«} \; ln[\frac{\boldsymbol{R}}{\bar{\boldsymbol{R}}}] \right]$$

we immediately see that the quantum potential is a four-dimensional scalar and hence invariant under Lorentz transformations.
This result enforces the hypothesis that any measurable quantum non-local process (even involving a large distance) is compatible with the postulate of invariance of light speed as the fastest way to which signals and interactions can be transmitted.
It is a matter of fact that the compatibility between the quantum mechanics and the postulate of light speed invariance of the relativity needs the definition of a theory able to describe the kinetic of the wave function collapse during the measurement process.
Actually, the dependence of the standard quantum theory from the measurement process makes it a semi empirical theory. On the other hand, a closed (self-standing) quantum theory must be able to describe the measuring process itself.
From the general point of view, the kinetics of an irreversible process (i.e., the measurement and/or wave function collapse) can be achieved with the help of the stochastic calculus applied to the quantum motion equations. To this end the QHA shows to be a very suitable formalism [12, 27].

## 6. Conclusion

In the present paper, the coupled hydrodynamic-type quantum equations for the phase and the amplitude of the wave function of the relativistic Dirac equation have been derived.
The work shows that in the low speed limit the quantum hydrodynamic conservation equation of a charged particle with spin described by the Pauli equation is recovered. The hydrodynamic motion equations in the low velocity limit also lead to the correct expression of the quantum pseudo-potential for charged particles.
The output shows that it is possible to derive the relativistic form of the quantum potential responsible for the non-local behavior of quantum mechanics and for the coherent evolution of the superposition of states.
The Lorentz invariance of the quantum potential points out that the non-local quantum effects such as the correlation between far away quantum measurements do not violate the relativistic postulate of invariance of light speed as the fastest way to which signals and interactions can be transmitted.

### Nomenclature

° = square wave function modulus — number of particle $l^{-3}$
$m$ =mass particle — m
$\hbar$ = Plank's constant — m $l^2$ $t^{-1}$
c = light speed — l $t^{-1}$
$H$ = Hamiltonian of the system — m $l^2$ $t^{-2}$
$V_{qu}$= quantum potential energy — m $l^2$ $t^{-2}$

# Appendix A

Excluding the domains where $\Psi = 0$ (that for the regularity of the wave function solutions are of null volume) we can write

$$\frac{\partial}{\partial t} ln[\frac{\Psi}{\overline{\Psi}}] = \left( \frac{\partial}{\partial t} ln[\frac{\Psi_1}{\Psi_3 *}], \quad \frac{\partial}{\partial t} ln[\frac{\Psi_2}{\Psi_4 *}] \quad \frac{\partial}{\partial t} ln[\frac{\Psi_3}{\Psi_1 *}] \quad \frac{\partial}{\partial t} ln[\frac{\Psi_4}{\Psi_2 *}] \right)$$

$$= \frac{\partial}{\partial t} ln[\boldsymbol{R}] - \frac{\partial}{\partial t} ln[\boldsymbol{R}¥^{0}] + \frac{i}{\hbar} \left( \frac{\partial \boldsymbol{I}\left(\boldsymbol{S} + \boldsymbol{S}¥^{0}\right)}{\partial t} \right) \quad \text{(A.1)}$$

$$= \frac{\partial}{\partial t} ln[\frac{\boldsymbol{R}}{\boldsymbol{R}¥^{0}}] + \frac{i}{\hbar} \left( \frac{\partial \boldsymbol{I}\left(\boldsymbol{S} + \boldsymbol{S}¥^{0}\right)}{\partial t} \right)$$

# Appendix B

In detail we have

$$\frac{\hbar}{i}\frac{\partial}{\partial t} ln[\frac{\boldsymbol{R}}{\boldsymbol{R}¥^{0}}] + \left( \frac{\partial \boldsymbol{I}\left(\boldsymbol{S} + \boldsymbol{S}¥^{0}\right)}{\partial t} \right) = -\left( \left[\frac{1}{\Psi}\right] H_D \Psi + \left[\frac{1}{\overline{\Psi}}\right] H_D{}^{*} \overline{\Psi} \right)$$

$$= -\left( \left[\frac{1}{\Psi}\right]\left[ c¥^{0}¥^{i} \bullet \left( \frac{\hbar}{i}\nabla - eA \right)\right]\Psi + \left[\frac{1}{\overline{\Psi}}\right]\left[ c¥^{0}¥^{i} \bullet \left( -\frac{\hbar}{i}\nabla - eA \right)\right]\overline{\Psi} \right) - 2\left(¥^{0} mc^{2} + e¤\right)$$

$$= -\left( \left[\frac{1}{\Psi}\right]\left[ c¥^{0}¥^{i} \bullet \frac{\hbar}{i}\nabla \right]\Psi + \left[\frac{1}{\overline{\Psi}}\right]\left[ c¥^{0}¥^{i} \bullet -\frac{\hbar}{i}\nabla \right]\overline{\Psi} \right) \quad \text{(B.1)}$$

$$+ \left[\frac{1}{\Psi}\right]\left[ c¥^{0}¥^{i} \bullet eA \right]\Psi + \left[\frac{1}{\overline{\Psi}}\right]\left[ c¥^{0}¥^{i} \bullet eA \right]\overline{\Psi} - 2\left(¥^{0} mc^{2} + e¤\right)$$

$$= -\left( \left[ c¥^{0}¥^{i} \bullet \frac{\hbar}{i}\left[\frac{1}{\Psi}\right]\nabla \right]\Psi + \left[ c¥^{0}¥^{i} \bullet -\frac{\hbar}{i}\left[\frac{1}{\overline{\Psi}}\right]\nabla \right]\overline{\Psi} \right) + 2\left[ c¥^{0}¥^{i} \bullet eA \right] - 2\left(¥^{0} mc^{2} + e¤\right)$$

from where it follows that

$$\frac{\hbar}{i}\frac{\partial}{\partial t} ln[\frac{\boldsymbol{R}}{\overline{\boldsymbol{R}}}] + \left( \frac{\partial \boldsymbol{I}\left(\boldsymbol{S} + \overline{\boldsymbol{S}}\right)}{\partial t} \right)$$

$$= -\begin{pmatrix} \left[ c¥^{0}¥^{i} \bullet \nabla \right]\boldsymbol{I}\left(\boldsymbol{S} + \overline{\boldsymbol{S}}\right) + \left[ c¥^{0}¥^{i} \bullet \frac{\hbar}{i}\left[\frac{1}{\boldsymbol{R}}\right]\nabla \right]\boldsymbol{R} \\ -\left[ c¥^{0}¥^{i} \bullet \frac{\hbar}{i}\left[\frac{1}{\overline{\boldsymbol{R}}}\right]\nabla \right]\overline{\boldsymbol{R}} \end{pmatrix} + 2\left[ c¥^{0}¥^{i} \bullet eA \right] - 2\left(¥^{0} mc^{2} + e¤\right) \quad \text{(B.2)}$$

# Appendix C

From equation (29) it follows that

$$\frac{d\left(\boldsymbol{p}-eA\right)}{dt}=-e\left(\nabla ¤+\frac{\partial A}{\partial t}\right)-e\left(\dot{q}\times\nabla\times A\right)-2¥^{0}c^{2}\nabla m$$

$$-\frac{\hbar}{2i}\left[\dot{q}\bullet\left(\frac{1}{\boldsymbol{R}}\nabla\boldsymbol{R}-\frac{1}{\overline{\boldsymbol{R}}}\nabla\overline{\boldsymbol{R}}\right)+\frac{\partial}{\partial t}ln[\frac{\boldsymbol{R}}{\overline{\boldsymbol{R}}}]\right]+\nabla\dot{q}_i\left(\partial_i\boldsymbol{I}\left(\boldsymbol{S}+\overline{\boldsymbol{S}}\right)-eA_i\right) \quad \text{(C.1)}$$

$$=-e\left(\boldsymbol{E}+\dot{q}\times\boldsymbol{B}\right)-¥^{0}c^{2}\nabla m+\nabla\dot{q}_i\left(\partial_i\boldsymbol{I}\left(\boldsymbol{S}+\overline{\boldsymbol{S}}\right)-eA_i\right)$$

$$-\frac{\hbar}{2i}\nabla\left[\dot{q}\bullet\nabla ln[\frac{\boldsymbol{R}}{\overline{\boldsymbol{R}}}]+\frac{\partial}{\partial t}ln[\frac{\boldsymbol{R}}{\overline{\boldsymbol{R}}}]\right]$$

# Appendix D

Given the equation

$$i\hbar\frac{\partial}{\partial t}|\Psi_{\pm}|^2=\frac{1}{2m}\left(\Psi^*{}_{\pm}(\frac{\hbar}{i}\nabla-eA)^2\Psi_{\pm}-\Psi_{\pm}(\frac{\hbar}{i}\nabla-eA)^2\Psi^*{}_{\pm}\right)$$

$$=-\frac{\hbar^2}{2m}\nabla\left(\Psi^{\dagger}{}_{\pm}\nabla\Psi_{\pm}-\left(\nabla\Psi^{\dagger}\right)\Psi_{\pm\pm}\right)-2\frac{\hbar e}{2im}\nabla\bullet|\Psi_{\pm}|^2 A \quad \text{(D.1)}$$

considering just that one with the plus sign, we obtain

$$-\frac{\hbar^2}{2m}\nabla\left(\Psi^{\dagger}\nabla\Psi-\left(\nabla\Psi^{\dagger}\right)\Psi\right)-2\frac{\hbar e}{2im}\nabla\bullet|\Psi|^2 A$$

$$=-\frac{\hbar^2}{2m}\nabla\left(¡^{\dagger}\cdot{}^*\nabla(¡\cdot)-\left(\nabla(¡\cdot)^{\dagger}\right)(¡\cdot)\right)-2\frac{\hbar e}{2im}\nabla\bullet|\Psi|^2 A$$

$$=-\frac{\hbar^2}{2m}\nabla\left(¡^{\dagger}\cdot{}^*(¡\nabla\cdot+\cdot\nabla¡)-\left(\left(¡^{\dagger}\nabla\cdot{}^*+\cdot{}^*\nabla¡^{\dagger}\right)\right)(¡\cdot)\right)-2\frac{\hbar e}{2im}\nabla\bullet|\Psi|^2 A \quad \text{(D.2)}$$

$$=-\frac{\hbar^2}{2m}\nabla\left(\begin{array}{l}\left(¡¡^{\dagger}\cdot{}^*\nabla\cdot+\cdot\cdot{}^*¡^{\dagger}\nabla¡\right)\\-\left(\left(¡^{\dagger}¡\cdot\nabla\cdot{}^*+\cdot{}^*\cdot\left(\nabla¡^{\dagger}\right)¡\right)\right)\end{array}\right)-2\frac{\hbar e}{2im}\nabla\bullet|\Psi|^2 A$$

where it has been used the identity

$$|\Psi|=|\cdot|.$$

Moreover, from (D.2) it follows that

$$-\frac{\hbar^2}{2m}\nabla\left(\left(\cdot{}^*\nabla\cdot-\cdot\nabla\cdot{}^*\right)+|\cdot|^2\left(¡^{\dagger}\nabla¡-\left(\nabla¡^{\dagger}\right)¡\right)\right)-2\frac{\hbar e}{2im}\nabla\bullet|\Psi|^2 A$$

$$=-\frac{\hbar^2}{2m}\nabla\left(\left(|\cdot|^2\nabla ln\frac{\cdot}{\cdot{}^*}\right)+|\cdot|^2\left(¡^{\dagger}\nabla¡-\left(\nabla¡^{\dagger}\right)¡\right)\right)-2\frac{\hbar e}{2im}\nabla\bullet|\cdot|^2 A \quad \text{(D.3)}$$

$$=-\frac{\hbar^2}{2m}\nabla\left(\left(|\cdot|^2\nabla ln\frac{\cdot}{\cdot{}^*}\right)-|\cdot|^2 i\left(cos^2\frac{g}{2}-sin^2\frac{g}{2}\right)\nabla¨\right)-2\frac{\hbar e}{2im}\nabla\bullet|\cdot|^2 A$$

and that

$$\frac{\partial}{\partial t}|\cdot\ |^2=-\nabla\bullet\left(|\cdot\ |^2\,\frac{1}{m}\nabla S\right)+\frac{1}{m}\nabla\bullet|\cdot\ |^2\,eA+\frac{\hbar}{2m}\nabla\bullet\left(|\cdot\ |^2\,cos\,\mathrm{g}\nabla^{\cdot\cdot}\right)$$

$$=-\nabla\bullet\left(|\cdot\ |^2\left(\frac{(\nabla S-eA)}{m}+cos\,\mathrm{g}\,\frac{\hbar}{2m}\nabla^{\cdot\cdot}\right)\right) \qquad \text{(D.4)}$$

# Appendix E

The total velocity time derivative and the total spin time derivative of the hydrodynamic representation of the Pauli equation can be obtained by the linear combinations of the system of two equations of (65). In fact, by taking the gradient of both members it follows that

$$\left((1,1)\frac{\partial\nabla S}{\partial t}+(-1,1)\frac{\hbar}{2}\frac{\partial\nabla^{\cdot\cdot}}{\partial t}\right)=$$

$$=-\frac{1}{2}(1,1)\nabla\begin{pmatrix}2e\text{¤}+\frac{(eA)^2}{m}-\frac{\hbar^2}{2m}\left(\left[\frac{1}{\Psi}\right]\nabla\bullet\nabla\Psi-\left[\frac{1}{\Psi^*}\right]\nabla\bullet\nabla\Psi^*\right)\\ -\frac{\hbar}{2im}\left(\left[\frac{1}{\Psi}\right]e(\nabla\bullet A\Psi)-\left[\frac{1}{\Psi^*}\right]e(\nabla\bullet A\Psi *)\right)\\ -\frac{\hbar}{2im}\left(\left[\frac{1}{\Psi}\right]eA\bullet\nabla\Psi-\left[\frac{1}{\Psi^*}\right]eA\bullet\nabla\Psi^*\right)\end{pmatrix} \qquad \text{(E.1)}$$

$$-\frac{1}{2}\nabla\left(\left[\frac{1}{\Psi}\right](\text{«}\boldsymbol{\sigma}\bullet B)\Psi+\left[\frac{1}{\Psi^*}\right](\text{«}\boldsymbol{\sigma}\bullet B)\Psi^*\right)$$

Moreover, by multiplying both members of the above equation by the matrix $\begin{bmatrix}cos^2\frac{\mathrm{g}}{2} & 0\\ 0 & sin^2\frac{\mathrm{g}}{2}\end{bmatrix}$ and then by making the summation of the two equations it follows that

$$\left( (sin^2 \frac{g}{2}, cos^2 \frac{g}{2}) \frac{\partial \nabla S}{\partial t} + (sin^2 \frac{g}{2}, -cos^2 \frac{g}{2}) \frac{\hbar}{2} \frac{\partial \nabla \text{¨}}{\partial t} \right) =$$

$$= -\frac{1}{2}(sin^2 \frac{g}{2}, cos^2 \frac{g}{2})\nabla \begin{pmatrix} 2e\text{¤} + \frac{(eA)^2}{m} - \frac{\hbar^2}{2m}\left( \left[ \frac{1}{\Psi} \right] \nabla \bullet \nabla \Psi - \left[ \frac{1}{\Psi^*} \right] \nabla \bullet \nabla \Psi^* \right) \\ -\frac{\hbar}{2im}\left( \left[ \frac{1}{\Psi} \right] e(\nabla \bullet A\Psi) - \left[ \frac{1}{\Psi^*} \right] e(\nabla \bullet A\Psi *) \right) \\ -\frac{\hbar}{2im}\left( \left[ \frac{1}{\Psi} \right] eA \bullet \nabla \Psi - \left[ \frac{1}{\Psi^*} \right] eA \bullet \nabla \Psi^* \right) \end{pmatrix} \quad \text{(E.2)}$$

$$-\frac{1}{2}\begin{bmatrix} cos^2 \frac{g}{2} & 0 \\ 0 & sin^2 \frac{g}{2} \end{bmatrix} \nabla \left( \left[ \frac{1}{\Psi} \right] (\text{«}\boldsymbol{\sigma} \bullet B)\Psi + \left[ \frac{1}{\Psi^*} \right] (\text{«}\boldsymbol{\sigma} \bullet B)\Psi^* \right)$$

and that

$$\left( \frac{\partial \nabla S}{\partial t} - (cos^2 \frac{g}{2} - sin^2 \frac{g}{2}) \frac{\hbar}{2} \frac{\partial \nabla \text{¨}}{\partial t} \right) =$$

$$= -\frac{1}{2}\nabla \begin{pmatrix} 2e\text{¤} + \frac{(eA)^2}{m} - \frac{\hbar^2}{2m}\left( \left[ \frac{1}{\Psi} \right] \nabla \bullet \nabla \Psi - \left[ \frac{1}{\Psi^*} \right] \nabla \bullet \nabla \Psi^* \right) \\ -\frac{\hbar}{2im}\left( \left[ \frac{1}{\Psi} \right] e(\nabla \bullet A\Psi) - \left[ \frac{1}{\Psi^*} \right] e(\nabla \bullet A\Psi *) \right) \\ -\frac{\hbar}{2im}\left( \left[ \frac{1}{\Psi} \right] eA \bullet \nabla \Psi - \left[ \frac{1}{\Psi^*} \right] eA \bullet \nabla \Psi^* \right) \end{pmatrix} \quad \text{(E.3)}$$

$$-\frac{1}{2}\begin{bmatrix} cos^2 \frac{g}{2} & 0 \\ 0 & sin^2 \frac{g}{2} \end{bmatrix} \nabla \left( \left[ \frac{1}{\Psi} \right] (\text{«}\boldsymbol{\sigma} \bullet B)\Psi + \left[ \frac{1}{\Psi^*} \right] (\text{«}\boldsymbol{\sigma} \bullet B)\Psi^* \right) \bullet \begin{pmatrix} 1 \\ 1 \end{pmatrix}$$

that leads to

$$
\left(\frac{\partial\left(\nabla S-\frac{\hbar}{2}\cos \mathrm{g}\nabla\text{¨}\right)}{\partial t}\right)=
$$

$$
=-\frac{1}{2}\nabla\left(\begin{array}{l}
2e\text{¤}+\frac{(eA)^2}{m}-\frac{\hbar^2}{2m}\left(\left[\frac{1}{\Psi}\right]\nabla\bullet\nabla\Psi-\left[\frac{1}{\Psi^*}\right]\nabla\bullet\nabla\Psi^*\right)\\
-\frac{\hbar}{2im}\left(\left[\frac{1}{\Psi}\right]e(\nabla\bullet A\Psi)-\left[\frac{1}{\Psi^*}\right]e(\nabla\bullet A\Psi *)\right)\\
-\frac{\hbar}{2im}\left(\left[\frac{1}{\Psi}\right]eA\bullet\nabla\Psi-\left[\frac{1}{\Psi^*}\right]eA\bullet\nabla\Psi^*\right)
\end{array}\right) \tag{E.3}
$$

$$
-\frac{1}{2}\begin{bmatrix}\cos^2\frac{\mathrm{g}}{2} & 0\\ 0 & \sin^2\frac{\mathrm{g}}{2}\end{bmatrix}\nabla\left(\left[\frac{1}{\Psi}\right](\text{«}\boldsymbol{\sigma}\bullet B)\Psi+\left[\frac{1}{\Psi^*}\right](\text{«}\boldsymbol{\sigma}\bullet B)\Psi^*\right)\bullet\begin{pmatrix}1\\1\end{pmatrix}
$$

and so on.